\documentclass[12pt,a4paper]{article}
\usepackage{latexsym,graphicx}
\newcommand{\eqref}[1]{(\ref{#1})}

\newcommand{\be}{\begin{equation}}
\newcommand{\ee}{\end{equation}}
\newcommand{\bea}{\begin{eqnarray}}
\newcommand{\eea}{\end{eqnarray}}

\begin{document}
\title{Invariant formulation of the Functional Renormalisation Group method for $U(n)\times U(n)$ symmetric matrix models }
\author{A. Patk\'os\\
Institute of Physics, E\"otv\"os University\\
H-1117, P\'azm\'any P\'eter s\'et\'any 1/A, Budapest, Hungary}
\vfill
\maketitle
\begin{abstract}
The Local Potential Approximation (LPA) to the Wetterich-equation is formulated explicitly in terms of operators, which are invariant under the $U(n)\times U(n)$ symmetry group. Complete formulas are presented for the two-flavor ($U(2)\times U(2)$) case. The same approach leads to a unique natural truncation of the functional driving the renormalisation flow of the potential of the three-flavor case ($U(3)\times U(3)$). The procedure applied to the $SU(3)\times SU(3)$ symmetric theory, results in an equation, which potentially allows an RG-investigation of the effect of the 't Hooft term representing the $U_A(1)$ anomaly, disentangled from the other operators.
\end{abstract}
\section{Introduction and General Analysis}
The $U_L(N)\times U_R(n)$ chiral symmetry is relevant to the QCD of $n$ massless quark flavors. Quark masses below $\Lambda_{QCD}$ introduce tolerable explicit symmetry breaking
which leaves this symmetry still a valid approximate concept underlying the construction of effective low energy models of strong interactions. At zero temperature 
the chiral symmetry is realized in the broken phase. The breakdown originates from two sources. It is spontanously broken to $U_V(n)$ \cite{nambu60,gell-mann60,nambu61} and the axial $U_A(1)$ quantum anomaly \cite{adler69,bell69} further breaks it to $U_V(1)\times SU_V(n)$. Low energy effective meson models are constructed along this specific breakdown of chiral symmetry.

Both the strength of the axial anomaly and the chiral condensate diminish with increasing temperature and eventually the full $U_L(n)\times U_R(n)$ symmetry will be restored.
It is an interesting issue to understand how wide is the domain of attraction of the symmetry breaking pattern observed in nature. Is there any sign of fine tuning in the
observed spectra of light mesons? More specifically, one should ask if a slight change of the couplings in the UV (high temperature) regime could result in a qualitatively different IR (low temperature) ground state. 
For instance, for $n=2$ there
are two alternative breaking patterns: $U_L(2)\times U_R(2)\rightarrow U_V(2)\rightarrow U_V(1)\times SU(2)$ and $U_L(2)\times U_R(2)\rightarrow U(1)\times U(1)\times U(1)$. Similarly, for $n=3$
there exists the possibility for a chain of symmetry breaking evolution: $U_L(3)\times U_R(3)\rightarrow U_V(3)\rightarrow U_V(2)\times U(1)$ even if one does not consider 
the effect of the axial anomaly. This alternative scenario was investigated recently with help of a large-$N$ approximation\cite{fejos11,fejos12}.

A most adequate tool to answer this kind of questions is provided by the Wilsonian Renormalisation Group analysis \cite{wilson74} of the scale evolution of the effective potential (or free energy at finite temperature). Starting with some initial potential at small distances the potential settles in one of the minima of the large wavelength 
effective potential. A global exploration of the dependence of the symmetry breaking pattern realised in the absolute minimum on the initial set of couplings decides on how robust is the actual scheme. A non-perturbative realisation of this program is offered by the various formulations of the Functional
 Renormalisation Group (FRG) equations \cite{wegner72,polchinski84,wetterich91,polonyi01}. 

Several studies in the $n=2$ case were done concentrating on the symmetry breaking pattern which is the closest 
to what is observed in nature \cite{berges97,fuku11,jiang12}. But the full landscape was not explored yet, therefore we do not know how generic is this pattern. 
The effect of the
anomaly was not included into the investigations, though this becomes particularly interesting in the $\eta-\eta^\prime$ phenomenology, present only for $n=3$.

There is a large number of FRG-investigations of the two-flavor effective models where the full symmetry is restricted to $SU(2)\times SU(2)\sim O(4)$\cite{bohr01,schaefer05,stokic10}, and the symmetry breaking pattern $O(4)\rightarrow O(3)$ is investigated also by taking into account confinement effects \cite{herbst11}. In the context of cold quantum gases also the $O(4)\rightarrow O(2)\times O(2)$ breakdown was studied with FRG approach\cite{andersen11}.
 
This note gives some technical details necessary to the full exploration of the coupling space for the $U(n)\times U(n)$ symmetric models, $n=2,3$. We present the most convenient form of the RG-flow equations which will be formulated explicitly in terms of variables invariant under the symmetry and respecting the analiticity features of the potential in these variables.  
	
In this investigation we use the so-called Local Potential Aproximation of the Wetterich equation \cite{berges02} 
\begin{equation}
\frac{\partial U_k(\varphi)}{\partial k}=\frac{1}{2}{\textrm Tr}\int_q \frac{\partial R_k(q)}{\partial k}\delta_{ij}\left[q^2+R_k(q)+\frac{\partial
^2U_k}{\partial \varphi_1\partial\varphi_2}\right]^{-1}_{ji},
\end{equation}
with a specific choice of the cut-off function suppressing low momentum modes in the functional integration \cite{litim01}
\begin{equation}
R_k(q)=(k^2-q^2)\Theta(k^2-q^2).
\end{equation}
It leads to the equation 
\begin{equation}
\partial_kU_k=\frac{k^{d+1}}{d}\frac{\Omega_d}{(2\pi)^d}\sum_{eigen-modes,\alpha}\frac{g_\alpha}{k^2+M_\alpha^2}=\frac{k^{d+1}}{d}\frac{\Omega_d}{(2\pi)^d}\frac{d}{dk^2}\log{\bf \Pi}[(k^2+M^2_\alpha)^{g_\alpha}],
\label{litim-cutoff}
\end{equation}
where $d$ is the dimension of the space-time, $\Omega_d$ the surface of the $d$-dimensional unit sphere and the sum $\alpha$ runs over the different eigenmodes of the mass-square matrix with eigenvalues $M_\alpha^2$ of degeneracy $g_\alpha$. The second equality makes it explicit that the scale evolution of the potential is driven by the characteristic polynomial ${\bf \Pi}[(k^2+M^2_\alpha)^{g_\alpha}]$ of the mass-square matrix.

The potential $U_k$ is invariant under the symmetry transformations, therefore it depends only on invariant combinations of the field variables. In the course of the derivation of the Wetterich equation one assumes that it has a power expansion in terms of these variables. Also the right hand side of (\ref{litim-cutoff}) should share this feature. We refer to this requirement below as the analyticity of the right hand side of (\ref{litim-cutoff}). 

The matrix field of the $U(n)\times U(n)$ invariant theory is parametrized with help of the generalized Gell-Mann matrices as
\begin{equation}
\Phi=\phi_l\lambda^l,\qquad \phi_l=s_l+i\pi_l, \quad l=0,1,2,...,n^2-1,\qquad {\textrm {Tr}}\lambda^l\lambda^k=2\delta_{kl}.
\end{equation}
The quadratic and quartic $U(n)\times U(n)$ invariants are defined as
\begin{equation}
\rho\equiv {\textrm {Tr}}\Phi^\dagger\Phi=2(s_l^2+\pi_l^2),\qquad
 \tau\equiv{\textrm {Tr}}\left(\Phi^\dagger\Phi-\frac{1}{n}{\textrm Tr\Phi^\dagger\Phi}\right)^2.
\label{invariants}
\end{equation}
Usually, one considers only $U(\rho,\tau)$, since these are the two operators, perturbatively renormalisable around the Gaussian fixed point. The expression of the quadratic invariant in terms of the quadratic fields is universal for all $n$ values. Note, that already for $n=3$ one has also higher power group invariants. 

The mass-square matrix is evaluated in terms of the derivatives of the invariants with respect to the field components and the derivatives of the potential
with respect to the invariants. For a specific parametrisation one chooses nonzero values for certain $s_i$ variables ($s_l=v_l, \pi_l=0$). On this background there is no mixing between the $s$- and the $\pi$-modes. The genuine expression is identical for both, therefore we write the explicit form only for the  $s$-sector: 
\begin{eqnarray}
_sM_{ij}^2&\equiv&\frac{\partial^2U_k}{\partial s_i\partial s_j}_{|_{s_i=v_i}}
=\frac{\partial^2 \rho}{\partial s_i\partial s_j}\frac{\partial U_k}{\partial\rho}
+\frac{\partial^2 \tau}{\partial s_i\partial s_j}\frac{\partial U_k}{\partial\tau}\nonumber\\
&+&\frac{\partial \rho}{\partial s_i}\frac{\partial \rho}{\partial s_j}
\frac{\partial^2 U_k}{\partial\rho^2}+\frac{\partial \tau}{\partial s_i}\frac{\partial \tau}{\partial s_j}
\frac{\partial^2 U_k}{\partial\tau^2}
+\left[\frac{\partial \rho}{\partial s_i}\frac{\partial \tau}{\partial s_j}+\frac{\partial \rho}{\partial s_j}\frac{\partial \tau}{\partial s_i}\right]
\frac{\partial^2U_k}{\partial \rho\partial\tau}.
\label{mass-matrix}
\end{eqnarray}
The derivatives of $\rho$ are the same for all $n$:
\begin{eqnarray}
&
\displaystyle
\frac{\partial \rho}{\partial s_i}=4s_i\rightarrow 4v_i,\qquad \frac{\partial \rho}{\partial \pi_i}=4\pi_i\rightarrow 0,
\nonumber\\
&\displaystyle
\frac{\partial^2 \rho}{\partial s_i\partial s_j}=4\delta_{ij}=\frac{\partial^2 \rho}{\partial \pi_i\partial \pi_j}.
\label{rho-derivative}
\end{eqnarray}
From here one can see that the first term on the right hand side of (\ref{mass-matrix}) contributes to each mode and is field independent (equal to $4U_{k,\rho}$). In the indices the quantity appearing after the prime 
refers to the partial derivative of the potential with respect to the corresponding invariant. Then one can write the factors $k^2+M_\alpha^2$ appearing in (\ref{litim-cutoff}) as $K^2+\mu_\alpha^2,~K^2=k^2+4U_{k,\rho}$, where $\mu^2_\alpha$ is the remainder of the mass eigenvalue after subtracting from it the field independent piece. In the $U(n)\times U(n)$ symmetric theory the $v$-dependent coefficients of the derivatives of $U_k$ start quadratically.
The expansion of the characteristic polynomial in powers of $K^2$ then starts as:
\begin{equation}
\Pi_\alpha(K^2+\mu_\alpha^2)^{g_\alpha}=(K^2)^N+(K^2)^{N-1}{\textrm{Tr}}\mu^2+1/2\cdot (K^2)^{N-2}(({\textrm{Tr}}\mu^2)^2-{\textrm{Tr}}(\mu^2)^2)+{\cal O}(v^6),
\label{K2-power-expansion}
\end{equation}
which after throwing away terms which contain in the prefactor of the derivatives of $U_k$ at least six powers of the $v$-background, leads on the right hand side of (\ref{litim-cutoff}) to the ratio
\begin{eqnarray}
\label{right-hand-side}
&&\sum_{eigen-\alpha}\frac{g_\alpha}{k^2+4U_{k,\rho}+\mu^2_\alpha}\approx\\
&&\frac{N(K^2)^{N-1}+(N-1)(K^2)^{N-2}{\textrm{Tr}}\mu^2+(N-2)/2\cdot (K^2)^{N-3}(({\textrm{Tr}}\mu^2)^2-
{\textrm{Tr}}(\mu^2)^2)}
{(K^2)^N+(K^2)^{N-1}{\textrm{Tr}}\mu^2+1/2\cdot (K^2)^{N-2}(({\textrm{Tr}}\mu^2)^2-{\textrm{Tr}}(\mu^2)^2)},\nonumber
\end{eqnarray}
($N=\sum g_\alpha=2n^2$). For the derivation of the present form one has to recognize that
\begin{equation}
 \sum_{eigen-\alpha}g_\alpha\mu^2_\alpha={\textrm {Tr}} \mu^2,\qquad  \sum_{eigen-\alpha}g_\alpha\mu^4_\alpha={\textrm {Tr}} (\mu^2)^2.
\end{equation}
The naturalness condition for this truncation will be clarified in the section discussing the 3-flavor model.

Assuming further $||\mu^2||\ll K^2$ one arrives at the form which is used to derive the leading order perturbative RG-running of the couplings characterising the lowest power operators figuring in the power series expansion of $U_k$\cite{pisarski84}:
\begin{equation}
\partial_kU_k=\frac{k^{d+1}}{d}\frac{\Omega_k}{(2\pi)^d}\sum_{eigen-\alpha}\left(\frac{g_\alpha}{K^2}-\frac{g_\alpha\mu^2_\alpha}{(K^4)^2}+\frac{g_\alpha (\mu_\alpha)^4}
{(K^6)^3}+...\right).
\label{pert-exp}
\end{equation} 
One should note, however, that normally one integrates (\ref{litim-cutoff}) down to $k=0$ and in some regions of the field variables (especially near the minimum of the potential) the assumed inequality might not hold. This circumstance restricts the range where one can rely on solutions of the Wetterich equation in terms of a power series.

In Section 2 the complete expression will be given for $n=2$, e.g. without any truncation in the characteristic polynomial. In Section 3 we argue that the omitted group invariant makes it inconsistent to keep any term of ${\cal O}(v^6)$ in the expansion of the characteristic plynomial, which yields a natural criterium on how and where to truncate. In Section 4 we consider the same truncation in case of $SU(3)\times SU(3)$ symmetry, where the cubic invariant $({\textrm {det}}\Phi+{\textrm {det}}\Phi^\dagger)/2$ will be introduced. The explicit formulae appearing in these three sections might be known to some workers actively pursuing FRG-investigations of the linear sigma model, but to our best knowledge they did not appear in previous publications. A short summary and outlook closes this note.

\section{Complete LPA equation for the $U(2)\times U(2)$ symmetric model}

There is no further invariant beyond those in (\ref{invariants}) in this case. We choose for the background non-zero $v_0$ and $v_3$. The quartic invariant looks then explicitly as
\begin{equation}
\tau=16\left[(s_0s_l+\pi_0\pi_l)^2+\pi_l^2s_k^2-(\pi_ls_l)^2\right]\rightarrow 16v_0^2v_3^2,\qquad k,l=1,2,3.
\end{equation}
After subsituting the non-zero field values its derivatives have the following expressions:
\begin{eqnarray}
&\displaystyle
\frac{\partial\tau}{\partial s_0}=32(s_0s_k+\pi_0\pi_k)s_k\rightarrow 32v_0v_3^2,\qquad 
\frac{\partial\tau}{\partial \pi_0}=32(s_0s_k+\pi_0\pi_k)\pi_k\rightarrow 0\nonumber\\
&\displaystyle
\frac{\partial\tau}{\partial s_l}=32[(s_0s_l+\pi_0\pi_l)s_0+\pi_k^2s_l-2\pi_ks_k\pi_l]\rightarrow 32v_0^2v_3\delta_{l3},
\nonumber\\
&\displaystyle
\frac{\partial\tau}{\partial \pi_l}=32[(s_0s_l+\pi_0\pi_l)\pi_0+s_k^2\pi_l-2\pi_ks_ks_l]\rightarrow 0,\nonumber\\
&\displaystyle
\frac{\partial^2\tau}{\partial\pi_0\partial\pi_0}=32\pi_k^2\rightarrow 0,\qquad \frac{\partial^2\tau}{\partial\pi_0\partial\pi_l}=32v_0v_3\delta_{l3},\qquad \frac{\partial^2\tau}{\partial\pi_l\partial\pi_k}=32v_3^2\delta_{lj},~~l\neq 3,\nonumber\\
&\displaystyle
\frac{\partial^2\tau}{\partial s_0\partial s_0}=32v_3^2,\qquad \frac{\partial^2\tau}{\partial s_0\partial s_l}=64v_0v_3\delta_{l3},\qquad \frac{\partial^2\tau}{\partial s_l\partial s_k}=32v_0^2\delta_{lj}.
\end{eqnarray} 
The mass matrix is built up readily and one finds directly two degenerate eigenvalues ($l=1,2$) plus the coupled $(0,3)$ modes both in the scalar and the pseudoscalar sectors. In the matrix eigenvalues one can replace $v_0$ and $v_3$ by the two invariants, but the individual matrix elements will show rather non-analytic dependence on $\tau$ and $\rho$ (see for instance Eqs.(5.2) and (5.3) of Ref.\cite{berges97}). Any truncation consisting of omitting some of the modes requires particular care in order not to loose the analiticity forced upon by the construction of the equation.

There is no need here to form the full characteristic polynomial of the mass matrix, since after some trials one finds that the contribution of the modes grouped into appropriate pairs can be expressed in polynomial form through $\rho$ and $\tau$. Actually, for each of these two-dimensional subspaces the three terms appearing in (\ref{K2-power-expansion}) represent the exact result (there is no ${\cal O}(v^6)$ contribution). In the following formulae we denote the scalar eigenmodes with an extra index $s$, and the pseudoscalars with $p$. One finds:
\begin{eqnarray}
(K^2+\mu_{s11}^2)(K^2+\mu_{p11}^2)&=&(K^2+\mu_{s22}^2)(K^2+\mu_{p22}^2)=K^4+AK^2+B, \nonumber\\
A=16\rho U_{k,\tau},&\qquad& B=64\tau U_{k,\tau}^2\nonumber\\
(K^2+\mu_{p03+}^2)(K^2+\mu_{p03-}^2)&=&K^4-64\tau U_{k,\tau}^2\nonumber\\
(K^2+\mu_{s03+}^2)(K^2+\mu_{s03-}^2)&=&K^4+CK^2+D,\nonumber
\end{eqnarray}
\begin{eqnarray}
C=&8&\left[2\rho U_{k,\tau}+\rho U_{k,\rho\rho}+4\rho\tau U_{k,\tau\tau}+4\tau U_{k,\tau\rho}\right],\nonumber\\
D=&64&\Bigl[4\tau(\tau-\rho^2)(U_{k,\rho\tau}^2-U_{k,\rho\rho}U_{k,\tau\tau})\nonumber\\
&-&U_{k,\tau}\left(3\tau U_{k,\tau}+6\tau^2U_{k,\tau\tau}+4\rho\tau U_{k,\rho\tau}+
(3\tau-2\rho^2)U_{k,\rho\rho}\right)\Bigr].
\label{characteristic-function-2}
\end{eqnarray}
Here the two eigenvalues emerging from the coupled pseudoscalar (0,3) modes are indexed as $\mu^2_{p03\pm}$ and for the scalar sector the corresponding notation $\mu^2_{s03\pm}$ is used.  

The pieces in (\ref{characteristic-function-2}) are substituted into (\ref{litim-cutoff}) and one arrives at the full LPA equation.  The exact invariant representation of the right hand side of this equation in the $U(2)\times U(2)$ case is analytic in the sense that a power series expansion in integer powers of $\rho$ and $\tau$ obviously exists.

It is clear that keeping only the lightest scalar and three of the pseudscalars, which would reduce this model to the usual pion-sigma model, leads to an inadvertent violation of the analiticity of the potential in $\rho$ and $\tau$.  

Phenomenological informations suggest that the chiral symmetry breaking pattern does not imply any breakdown of the isospin symmetry, e.g. the minimum of the effective potential is expected to be located at small values of $\tau$  (corresponding in our parametrisation to small $v_3$), though its quantum fluctuations contribute to the RG-evolution of the potential. This remark suggests the simplifying ansatz \cite{berges97,fuku11}:
\begin{equation}
U(\rho,\tau)=U^{(1)}(\rho)+U^{(2)}(\rho)\tau,
\end{equation}
replacing the problem of the determination of a two-variable function by two one-variable ones. Technically, one substitutes this ansatz and its appropriate derivatives into (\ref{characteristic-function-2}) and expands the expression of the right hand side of (\ref{litim-cutoff}) in the variable $\tau$ up terms linear in $\tau$. By comparing the coefficients of $\tau^0$ and of $\tau^1$ one finds the one variable equations solved in Refs.\cite{berges97,fuku11} without the need for any extra consideration concerning the cancellation of apparent singular contributions unavoidably showing up when this expression is written in terms of $v_0$ and $v_3$. With a further approximation it might be convenient for phenomenological purposes to fix the potential to the perturbatively renormalisable form \cite{chan73,lenaghan00} and use Eq.(\ref{pert-exp}) just to trace the scale dependence of its couplings\cite{jiang12}.  It would be, however interesting to see if there is in the potential any non-trivial structure along the $\tau$ direction, which can be studied by keeping the full expression on the right hand side of (\ref{litim-cutoff}). 

\section{Consistent truncation in the $U(3)\times U(3)$ case}

The pattern of chiral symmetry breaking in the three-flavor case is mostly investigated in the $v_0-v_8$-plane. We also used this background for the parametrisation of the quadratic and quartic invariants:
\begin{eqnarray}
\rho&=&2(s_a^2+\pi_a^2)\rightarrow 2(v_0^2+v_8^2)\nonumber\\
\tau&=&8\left[\left(\frac{1}{2}d_{ijk}(s_is_j+\pi_i\pi_j)+\sqrt{\frac{2}{3}}(s_0s_k+\pi_0\pi_k)\right)^2
+(f_{ijk}\pi_is_j)^2\right]\nonumber\\
&\rightarrow& \frac{2}{3}v_8^2(v_8-2\sqrt{2}v_0)^2.
\end{eqnarray}
The derivatives necessary for building up the mass-square matrix are simply obtained for $\rho$, and after a lengthy, though straightforward calculation one finds the following formulae for the derivatives of $\tau$:
\begin{eqnarray}
\frac{\partial\tau}{\partial s_l}&=&\frac{8}{3}v_8(\sqrt{2}v_0-v_8)(2\sqrt{2}v_0-v_8)\delta_{l8},\qquad l=1,...,8\nonumber\\
\frac{\partial\tau}{\partial s_0}&=&\frac{8\sqrt{2}}{3}v_8^2(2
\sqrt{2}v_0-v_8),\qquad \frac{\partial\tau}{\partial\pi^a}=0,
\end{eqnarray}
\begin{eqnarray}
\frac{\partial^2\tau}{\partial s_k\partial s_l}&=&16\delta_{kl}\left[\left(d_{8ll}v_8+\sqrt{\frac{2}{3}}v_0\right)^2+d_{8ll}v_8\left(-\frac{1}{2\sqrt{3}}v_8+\sqrt{\frac{2}{3}}v_0\right)\right],\nonumber\\
\frac{\partial^2\tau}{\partial s_0\partial s_l}&=&\frac{32}{3}v_8\delta_{l8}\left[2v_0-\frac{3}{2\sqrt{2}}v_8\right],\nonumber\\
\frac{\partial^2\tau}{\partial s_0^2}&=&\frac{32}{3}v_8^2,\qquad \frac{\partial^2\tau}{\partial\pi_0^2}=0,\nonumber\\
\frac{\partial^2\tau}{\partial\pi_k\partial\pi_l}&=&16\left[f_{k8n}f_{l8n}v_8^2+d_{8kl}\left(-\frac{1}{2\sqrt{3}}v_8^2+\sqrt{\frac{2}{3}}v_0v_8\right)\right],\nonumber\\
\frac{\partial^2\tau}{\partial \pi_0\partial \pi_l}&=&\frac{32}{3}v_8\delta_{l8}\left[v_0-\frac{1}{2\sqrt{2}}v_8\right].
\end{eqnarray}
Making use of the structural constants of the $U(3)$ group one constructs the mass-square matrix. Its scalar and pseudoscalar sectors are independent and of the same structure. The modes $k=1,...,7$ are directly found to be eigenmodes, and the modes $0,8$ again form two $2\times 2$  coupled subsystems. 

In the $U(3)\times U(3)$ case we could not identify smaller subsets of the factors of the characteristic polynomial which could be separately expressed through the invariants, therefore we turn to the expansion (\ref{K2-power-expansion}). For the terms displayed in this formula we find:
\begin{eqnarray}
{\textrm {Tr}} \mu^2&=&8\rho U_{k,\rho\rho}+\frac{128}{3}\rho U_{k,\tau}+32\tau U_{k,\rho\tau}+{\cal O}(v^6),\nonumber\\
{\textrm{Tr}}(\mu^2)^2&=&384\tau U_{k,\tau}U_{k,\rho\rho}+64\rho^2U^2_{k,\rho\rho}+\left(512\tau+\frac{2048}{9}\rho^2\right)\tau U^2_{k,\tau}+{\cal O}(v^6).
\label{traces}
\end{eqnarray}
In the first line the term ${\cal O}(v^6)$ refers to the coefficient of $U_{k,\tau\tau}$, which contains altogether six factors of $v_0$ and $v_8$. It receives contribution exclusively from the scalar $(0,8)$ sector:
\begin{equation}
\left(\frac{\partial\tau}{\partial s_0}\right)^2+\left(\frac{\partial\tau}{\partial s_8}\right)^2=\frac{32}{3}\tau\left(\rho+
{\textrm{sign}}[v_8^2-2\sqrt{2}v_0v_8]\sqrt{\frac{3\tau}{2}}\right).
\end{equation}
This contribution is clearly non-analytic, what is expected to appear just at this level since the $U(3)$ group has a further invariant ($\sim {\textrm{Tr}}(\Phi^\dagger\Phi)^3$) containing terms with six factors of the fields.
In the second line  of (\ref{traces}) among the omitted terms the lowest number of $v_i$-factors appears in front of $U_{k,\rho\tau}U_{k,\tau}$ and $U_{k,\rho\tau}U_{k,\rho\rho}$. Direct calculation demonstrates that none of these coefficients can be expressed through integer powers of the two invariants, i.e. in a form $c_1\rho^3+c_2\tau\rho$.
For this, also the group invariant containing six powers of the background should have been included.

The only consistent way of restricting the dependence of the potential to $\rho$ and $\tau$ and obey the analiticity is to truncate the expansion of the characteristic polynomial after the three first terms and solve the FRG-equation with (\ref{right-hand-side}) on its right hand side.

\section{Including the $U_A(1)$ anomaly: the $SU(3)\times SU(3)$ case}

An effective linear sigma model which reflects the $U_A(1)$ anomaly is phenomenologically more appealing. In phenomenological investigations\cite{chan73,lenaghan00} one does not separate from the matrix field $\Phi$ the zeroth component, which would produce a large number of operators in the FRG program. On the basis of processes mediated by instanton configurations one just adds to the $U(n)\times U(n)$ invariant Lagrangian a term which is less symmetric. The corresponding piece of the Lagrangian is 't Hooft's determinant: $\delta=({\textrm{det}}\Phi+{\textrm{c.c.}})/2$. It breaks the $U(3)\times U(3)$ symmetry down to $SU(3)\times SU(3)$. 

In terms of the parametrisation of the preceding section one has
\begin{equation}
\delta=\frac{1}{3\sqrt{3}}(\sqrt{2}v_0+v_8)^2(\sqrt{2}v_0-2v_8).
\end{equation}
Below we follow the usual process of taking into account the axial anomaly and just include $\delta$ into the set of variables of the effective potential. We consider $U(\rho,\delta,\tau)$ and calculate the complementary piece to the mass-square matrix of the model with help of the derivatives of $\delta$ with respect the scalar and pseudoscalar fields. These have the following form on the previous background:
\begin{eqnarray}
\frac{\partial\delta}{\partial s_i}&=&\frac{\partial\delta}{\partial \pi_a}=0,\qquad {\textrm {if}}~ i=1,...,7,\quad a=0,...,8,\nonumber\\
\frac{\partial\delta}{\partial s_0}&=&\sqrt{\frac{2}{3}}(2v_0^2-v_8^2),\qquad \frac{\partial\delta}{\partial s_8}=-\frac{2}{\sqrt{3}}v_8(v_8+\sqrt{2}v_0).
\end{eqnarray}
\begin{eqnarray}
\frac{\partial^2\delta}{\partial s_i\partial s_j}&=&-\frac{\partial^2\delta}{\partial \pi_i\partial \pi_j}=-2\sqrt{\frac{2}{3}}(v_0-\sqrt{2}v_8)\delta_{ij},\qquad {\textrm {if}}~ i=1,2,3,\nonumber\\
\frac{\partial^2\delta}{\partial s_i\partial s_j}&=&-\frac{\partial^2\delta}{\partial \pi_i\partial \pi_j}=-\frac{2}{\sqrt{3}}(\sqrt{2}v_0+v_8)\delta_{ij}, \qquad {\textrm{if}}~ i=4,5,6,7,\nonumber\\
\frac{\partial^2\delta}{\partial s_0^2}&=&-\frac{\partial^2\delta}{\partial \pi_0^2}=4\sqrt{\frac{2}{3}}v_0,
\qquad \frac{\partial^2\delta}{\partial s_8^2}=-\frac{\partial^2\delta}{\partial \pi_8^2}=-2\sqrt{\frac{2}{3}}(v_0+\sqrt{2}v_8),\nonumber\\
\frac{\partial^2\delta}{\partial s_0\partial s_8}&=&-\frac{\partial^2\delta}{\partial \pi_0\partial\pi_8}=-2\sqrt{\frac{2}{3}}v_8.
\end{eqnarray}
The opposite sign of the second derivatives with respect to the scalar and the pseudoscalar fields ensures the cancellation of the contributions to the mass-square matrix which would be linear in the background fields.

The completion of the mass-square matrix involving at least one partial derivative of $U_k$ with respect to $\delta$ is worked out and the following additions are found to the traces (\ref{traces}): 
\begin{eqnarray}
\Delta{\textrm{Tr}}\mu^2&=&\left(\frac{2}{3}\rho^2-\tau\right)U_{k,\delta\delta}+24\delta U_{k,\delta\rho}+{\cal O}(v^6),\nonumber\\
\Delta{\textrm{Tr}}(\mu^2)^2&=&32\rho U_{k,\delta}^2+192\delta U_{k,\rho\rho}U_{k,\delta}-1024\delta U_{k,\tau}U_{k,\delta}+{\cal O}(v^6).
\label{delta-traces}
\end{eqnarray}
One should wonder about the ${\cal O}(v^5)$ terms of these expressions.
In the first line it appears in front of $U_{k,\delta\tau}$, on the second line there is another ${\cal O}(v^5)$ coefficient in front of $U_{k,\rho\tau}U_{k,\delta}$.
Both are fully determined by the $(0,8)$ scalar sector having the respective coefficient:
\begin{eqnarray}
&\displaystyle
2\left(\frac{\partial\tau}{\partial s_0}\frac{\partial\delta}{\partial s_0}+\frac{\partial\tau}{\partial s_8}\frac{\partial\delta}{\partial s_8}\right),
\nonumber\\
&
\displaystyle
4\left[\frac{\partial\rho}{\partial s_0}\frac{\partial\tau}{\partial s_0}\frac{\partial^2\delta}{\partial s_0^2}+
\left(\frac{\partial\rho}{\partial s_0}\frac{\partial\tau}{\partial s_8}+\frac{\partial\rho}{\partial s_8}\frac{\partial\tau}{\partial s_0}\right)\frac{\partial^2\delta}{\partial s_0\partial s_8}+\frac{\partial\rho}{\partial s_8}\frac{\partial\tau}{\partial s_8}\frac{\partial^2\delta}{\partial s_8^2}\right].
\end{eqnarray}
  One verifies explicitly using the derivatives listed above that both coefficients vanish!
Therefore the truncation of the characteristic polynomial will be done again at terms containing no more than four factors in the background fields, though the accuracy is still ${\cal O}(v^6)$.

It is just the present system where the advantage of the invariant formulation of the RG-equation over the use of the component background becomes evident. Although taking appropriate derivatives of $U_k(v_0,v_8)$ one can trace the evolution of the coupling in front of $\delta$, with the $v_0-v_8$ component parametrisation one can explore only a surface in the three-dimensional $\rho-\tau-\delta$ space. On the other hand using the invariant formulation, e.g. solving Eqs. (\ref{litim-cutoff}) with (\ref{right-hand-side}), where the traces are given by the sum of (\ref{traces}) and (\ref{delta-traces}), one can explore the full space. Prospectively, by generalising the proposed procedure to finite temperature one can study also the thermal evolution of the expectation value of $\delta$, e.g. that of the $U_A(1)$ anomaly. It might give access to the old question whether the chiral symmetry or the axial anomaly is restored first\cite{pisarski84}.

The approach described in this note is followed often for $O(n)$ symmetric theories. There, however, a single invariant exists, therefore there is no big difference in using the invariant or the coordinate representation of the background. In case of the matrix models with unitary symmetry the formulation presented here allows a more transparent understanding of the symmetry features of the ground state. The choice of appropriate coordinates for the representation of the background is convenient to compute the coefficients of the derivatives of the effective potential. However, some cancellations (for instance, the vanishing of the coefficients of $U_{k,\delta\tau}$ and $U_{k,\rho\tau}U_{k,\delta}$) in presence of the determinant term appear somewhat miraculous when one uses explicit coordinate representation of the invariants, though it might have some simple explanation in terms of group theory. 

Numerical investigations should reveal the possible existence of local minima of the effective potential which correspond to symmetry breaking patterns beyond the pattern realized in nature.  The generalisation of the present treatment to the more general case of non-trivial field renormalisation is also worth of consideration.

\section*{Acknowledgments}
This research was supported by the Hungarian Research Fund Grants K77534. Useful discussions with G. Fej\H os, A. Jakov\'ac and Zs. Sz\'ep are gratefully acknowledged.

\end{document}